\documentclass[conference]{IEEEtran}
\IEEEoverridecommandlockouts

\usepackage{cite}
\usepackage{amsmath,amssymb,amsfonts}
\usepackage{algorithmic}
\usepackage{graphicx}
\usepackage{textcomp}
\usepackage{xcolor}
\usepackage{footmisc}
\usepackage{hyperref}
\def\BibTeX{{\rm B\kern-.05em{\sc i\kern-.025em b}\kern-.08em
    T\kern-.1667em\lower.7ex\hbox{E}\kern-.125emX}}

\begin{document}

\title{Q-BEAST: A Practical Course on Experimental Evaluation and Characterization of Quantum Computing Systems\\
}

\author{
\IEEEauthorblockN{
	Minh Chung$^{1,3}$, Yaknan Gambo$^{1,2}$, Burak Mete$^{1}$, Xiao-Ting Michelle To$^{3}$, Florian Krötz$^{3}$,\\
	Korbinian Staudacher$^{3}$, Martin Letras$^{1}$, Xiaolong Deng$^{1}$, Mounika Vavilala$^{1}$, Amir Raoofy$^{1}$,\\
    Jorge Echavarria$^{1}$, Luigi Iapichino$^{1}$, Laura Schulz$^{*}$, Josef Weidendorfer$^{1}$, Martin Schulz$^{1,2}$
    }
    \IEEEauthorblockA{$^{1}$Leibniz Supercomputing Centre (LRZ), Garching bei München, Germany}
    \IEEEauthorblockA{$^{2}$Chair for Computer Architecture and Parallel Systems, Technical University of Munich (TUM), Germany}
    \IEEEauthorblockA{$^{3}$MNM-Team, Ludwig-Maximilians-Universität München (LMU), Germany}
    \IEEEauthorblockA{
        \{minh.chung, yaknan.gambo, burak.mete, martin.letras, xiaolong.deng, mounika.vavilala, amir.raoofy,\\
        jorge.echavarria, luigi.iapichino, josef.weidendorfer\}@lrz.de, \{martin.w.j.schulz\}@tum.de,\\
        \{michelle.to, florian.kroetz, korbinian.staudacher\}@nm.ifi.lmu.de
    }
    \thanks{$^{*}$schulz@anl.gov, Argonne National Laboratory, Lemont, IL, USA}
}

\maketitle

\begin{abstract}
Quantum computing (QC) promises to be a transformative technology with impact on various application domains, such as optimization, cryptography, and material science. However, the technology has a sharp learning curve, and practical evaluation and characterization of quantum systems remains complex and challenging, particularly for students and newcomers from computer science to the field of quantum computing. To address this educational gap, we introduce Q-BEAST, a practical course designed to provide structured training in the experimental analysis of quantum computing systems. Q-BEAST offers a curriculum that combines foundational concepts in quantum computing with practical methodologies and use cases for benchmarking and performance evaluation on actual quantum systems. Through theoretical instruction and hands-on experimentation, students gain experience in assessing the advantages and limitations of real quantum technologies. With that, Q-BEAST supports the education of a future generation of quantum computing users and developers. Furthermore, it also explicitly promotes a deeper integration of High Performance Computing (HPC) and QC in research and education.
\end{abstract}


\section{Introduction}
Quantum computing (QC) has garnered attention due to its potential to solve a specific class of complex and important problems. However, to overcome the current limitation of QC, integrating High Performance Computing (HPC) with QC is a key motivation to maximize computational potential. Current quantum computers are still in the Noisy Intermediate-Scale Quantum (NISQ)~\cite{preskill2018nisq} era, in which we encounter instability, non-negligible error rates, and often a limited qubit count. Despite this, HPC, already at this state, allows QC workflows to exploit parallelism and enable efficient handling of large datasets and pre-/post-processing algorithms. While HPC manages the bulk of the workload, data preprocessing, and workflow controlling, QC tackles the particular tasks/segments where the quantum advantage is possible. 

Looking at the current state-of-the-art, various technologies and platforms are being developed, leading to a wide variety of quantum hardware modalities, such as superconducting~\cite{heimonen2023qciqm}, trapped ions~\cite{frisch2024qcaqt}, and photonics~\cite{slussarenko2019qcphotonic}. On the theoretical side, quantum algorithms have evolved rapidly. However, the practical side of mapping such algorithms to current hardware implementations remains complex and requires specialized knowledge. This motivates efforts to emphasize education on these concepts with new approaches and to evaluate and characterize QC systems.

Understanding the performance of quantum hardware is crucial to developing efficient quantum algorithms, and vice versa. In order to facilitate this understanding---and this not only in theory, but on practical and real examples---we created a practical course named Q-BEAST aimed at enabling a new generation of computer scientists to take advantage of this potentially revolutionary technology. The Q-BEAST framework is designed to address educational gaps by providing students with practical experience in benchmarking and evaluating quantum systems, allowing them to apply theoretical knowledge to real systems. Our curriculum includes the study of various aspects that significantly impact computational outcomes, including noise characteristics, gate fidelity, and crosstalk. Q-BEAST combines theoretical foundations with hands-on training, enabling students to implement effective measurement protocols, identify system bottlenecks, and evaluate the practical challenges in real-world quantum computing systems. The course structure emphasizes experimental best practices, with the following key topics:

\begin{itemize}
    \item Foundational Understanding: Overview of the background required to understand the HPC system, quantum computing, quantum hardware performance metrics, together with quantum algorithms.

    \item Experimental Methodologies: Guidance on experiments to evaluate system characteristics and compare with different simulators and real quantum devices.

    \item Benchmarking Techniques: Introduction to standardized benchmarking protocols that allow the comparison of different quantum systems.

    \item Hands-on Assignments and Projects: Engaging students in practical assignments and projects, on real on-premise quantum computing systems.

    \item Analysis and Feedback: Motivating students' abilities to analyze experimental data; after the course, as their feedback can help improve HPC-Quantum education efforts.
    
\end{itemize}

By integrating these topics into a structured curriculum, Q-BEAST offers students opportunities to learn about and access the emerging field of quantum computing. Alongside the focus on the experimental evaluation of quantum systems, Q-BEAST introduces students to the concept of hybrid High Performance Computing-Quantum Computing (HPCQC) workflows, which we see as central to enabling useful quantum computation now and moving forward. In the current NISQ era, quantum hardware still faces limitations in terms of fidelity or qubit scalability. Integrating HPC allows us to offload classical parts of hybrid quantum algorithms (e.g., optimization loops, warm-up states) and manage large-scale simulations efficiently. In the current setup of Q-BEAST, the course does not require the use of many compute nodes; however, the HPCQC concept provides an important foundation. It allows students to explore the coupling between classical parallelism and quantum execution.

The course is designed to be accessible to undergraduate and graduate students and the course's modular content can be adapted to various educational settings. This paper outlines the course structure, learning outcomes, and key experimental techniques used in Q-BEAST, offering a roadmap to enhance quantum computing education through practical evaluation and characterization methodologies.

\section{Motivation}

The motivation for developing the Q-BEAST lab course (so-called ``Praktikum'' in German) originates from three perspectives: the educational needs of the university, students' interests, and the research institution's goals.

\begin{itemize}
    \item \textbf{University Perspective:} From an educational standpoint, Q-BEAST offers an opportunity to bridge theoretical understanding with practical experiments. While traditional quantum computing courses focus heavily on theoretical concepts, this practical course addresses the crucial need for students to gain practical experience. Q-BEAST enables students to apply theoretical principles through realistic experimentation using on-premise quantum systems with all their capabilities and challenges, reinforcing their understanding of quantum systems and their ability to deal with real-life production scenarios.

    \item \textbf{Student Perspective:} Students are strongly motivated by the unique opportunity to access both classical/HPC resources and real quantum hardware. By participating in Q-BEAST, students gain practical experience in conducting experiments, analyzing data, and understanding the performance of the quantum system. This hands-on experience cultivates technical skills, encouraging curiosity and engagement while preparing students to contribute effectively to the evolving field of quantum computing.

    \item \textbf{Research Institution Perspective (LRZ):} At the Leibniz Supercomputing Centre (LRZ), we manage and explore different quantum computing modalities (technologies). Understanding the infrastructure and emerging paradigms of these new computing systems is crucial. Additionally, as part of our integration efforts, developing and enhancing the software stack that enables quantum computing serves as a crucial bridge between our QC and HPC systems. Through Q-BEAST, we can conduct systematic tests and gather valuable feedback to refine the requirements for this integration framework.
\end{itemize}

\section{HPCQC Infrastructure for the Lab Course}

The Q-BEAST infrastructure refers to the HPC and quantum computing resources available at the Leibniz Supercomputing Centre (LRZ). For our lab setup, we use a testbed system named BEAST\footnote{\url{https://www.lrz.de/technologien/future-computing}, where BEAST stands for Bavarian Energy, Architecture- and Software-Testbed.}, a specialized research platform that features different architectures, such as CPUs and GPUs from different hardware vendors like AMD, Intel, NVIDIA, etc. These diverse computing architectures enable researchers and students to explore the performance and complexity of different parallel computing architectures.

For the Q-BEAST Praktikum (lab course), we use a dedicated HPC cluster named Wolpertinger, which is part of the BEAST system. Wolpertinger is equipped with compute nodes, where each contains two-socket Intel CPUs (Intel Xeon Platinum 8360Y), one NVIDIA GPU A$100$ $80$GB, one Silcom FPGA SmartNIC, and $256$GB of main memory, providing sufficient resources to conduct experiments with quantum algorithms or hybrid HPC-quantum workflows.

\begin{figure}[t]
    \includegraphics[scale=0.75]{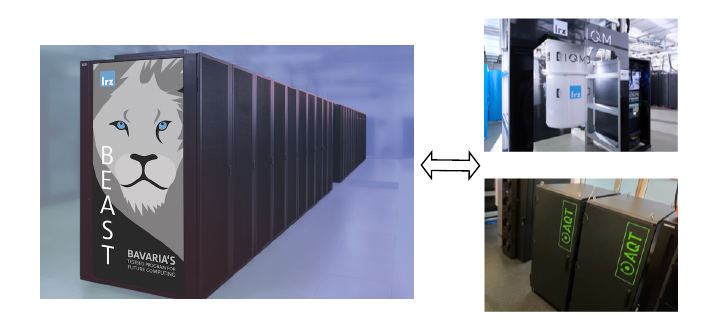}
    \caption{The BEAST testbed system and quantum computers at LRZ, the 20-qubit superconducting device (upper-right corner, codename QExa20) from IQM, and the 20-qubit ion-trapped device (lower-right corner, codename AQT20) from AQT.}
    \label{fig:beast_qc_systems}
\end{figure}

Wolpertinger showcases a co-located tight integration with quantum computers~\cite{elsharkawy2024hpcqcinteg}, enabling students to perform hybrid HPC-QC experiments. While large-scale parallelism is not a strict requirement for the assignments/projects in Q-BEAST, the inclusion of HPC is significant. It introduces students to the realities of scaling, resource management, and classical-quantum coupling components in the design of hybrid algorithms. This integration ensures that students gain practical experience deploying quantum applications on real systems. They can characterize system performance and explore potential optimizations for hybrid classical-quantum environments. Figure \ref{fig:beast_qc_systems} shows the infrastructure photo, where the BEAST system is illustrated with two of the available quantum computers at LRZ, the 20-qubits superconducting device from IQM~\cite{heimonen2023qciqm} and the 20-qubits ion-trapped device from AQT \cite{frisch2024qcaqt}. These devices are exposed to the software stack as backends with codenames QExa20, AQT20.\\

\subsection*{Munich Quantum Software Stack (MQSS)}

The software layer supporting HPCQC integration and access to quantum systems is referred to the Munich Quantum Software Stack (MQSS)~\cite{schulz2023mqss}. It is developed as part of the Munich Quantum Valley (MQV) initiative. MQSS is designed to provide a comprehensive and holistic software environment that facilitates direct access to quantum computers (so-called quantum backends). The stack supports seamless integration with HPC systems, enhanced quantum compilation, and optimization toolkits. There are two primary access pathways that we use to grant permission to students, as follows:

\begin{itemize}
    \item \textbf{Munich Quantum Portal}\footnote{\url{https://portal.quantum.lrz.de}} (MQP): A web-based interface that enables streamlined access to quantum devices. Each student has an account and can log into the portal. The MQP's dashboard shows all information regarding backend availability, job logs, and other details. To submit a job, users need to create an authorization token. At the user working environment, users need to install a Python package currently called \texttt{mqp-qiskit-provider}\footnote{At the time of writing, the current interface to MQSS for Qiskit is provided as \texttt{mqp-qiskit-provider}. As MQSS is under active development, provider interfaces' naming and modular structure may evolve. Therefore, in future iterations, it can be changed into a unified \texttt{mqss-adapter} supporting multiple providers, e.g., Qiskit, PennyLane, and even HPCQC-provider, to ensure flexibility and streamlined integration.\label{mqssnote}}, which has interfaces to load the token and access quantum backends at LRZ remotely~\cite{qis2025mqpqiskitprovider}.

    \item \textbf{HPCQC Access Path:} An integrated route via HPC systems, particularly BEAST and Wolpertinger, to connect with quantum hardware so that students can log into our system, prepare HPC jobs, QC jobs, or hybrid jobs, and submit them to run on real devices. Using the same given accounts, students can log into the BEAST system by using \texttt{SSH}~\cite{ylonen2006ssh}. Students prepare applications, job descriptions, and submit the jobs via the SLURM resource scheduler, known as a workload manager in HPC systems~\cite{yoo2003slurm}. Regarding quantum tasks in SLURM job submission, a similar package called HPCQC Offload Provider\footref{mqssnote} is used, which provides interfaces to offload quantum tasks to the quantum backends at LRZ. 
\end{itemize}

In addition, the software environment of BEAST and Wolpertinger is part of the LRZ research infrastructure, supporting active researchers from Munich universities, MQV, and internal LRZ users. Therefore, our software environment is subject to change in response to evolving research needs. For this reason, necessary HPC system packages/libraries such as Spack~\cite{gamblin2015spack}, Environment Modules~\cite{furlani1991envmod}, etc., are used to maintain a dedicated software for the practical course. Additionally, various compilers for different backends and tools are installed on BEAST as well as Wolpertinger.. We also deploy a system-wide instance called DataCenter DataBase (DCDB)~\cite{netti2019dcdb} to monitor the outbound power consumption of nodes gathered from Power Distribution Units (PDUs), which allows students to characterize the power information of quantum computers and HPC nodes.

\section{Q-BEAST Structure and Teaching Approach}

The Q-BEAST Praktikum focuses on emerging quantum computing architectures such as superconducting, ion-trapped, and neutral atom systems. The lab course covers a range of topics, including quantum computing fundamentals, system characterization, and performance evaluation. Assignments and projects require hands-on experiments spanning quantum simulation methods, quantum algorithms, benchmarking, qubit connectivity, topology awareness, and hybrid HPC-quantum use cases.

Students work in groups of two or three to complete assignments and projects. The assignment workload takes one or two weeks to complete, while the final project takes three weeks. Students need to split general questions, experiment with tasks, and work together to complete assignments by the deadline. Before conducting experiments, they are directed to develop hypotheses about the expected results. This can help them improve analytical skills and get deeper into quantum principles.

\begin{figure*}[t]
    \centering
    \includegraphics[scale=0.7]{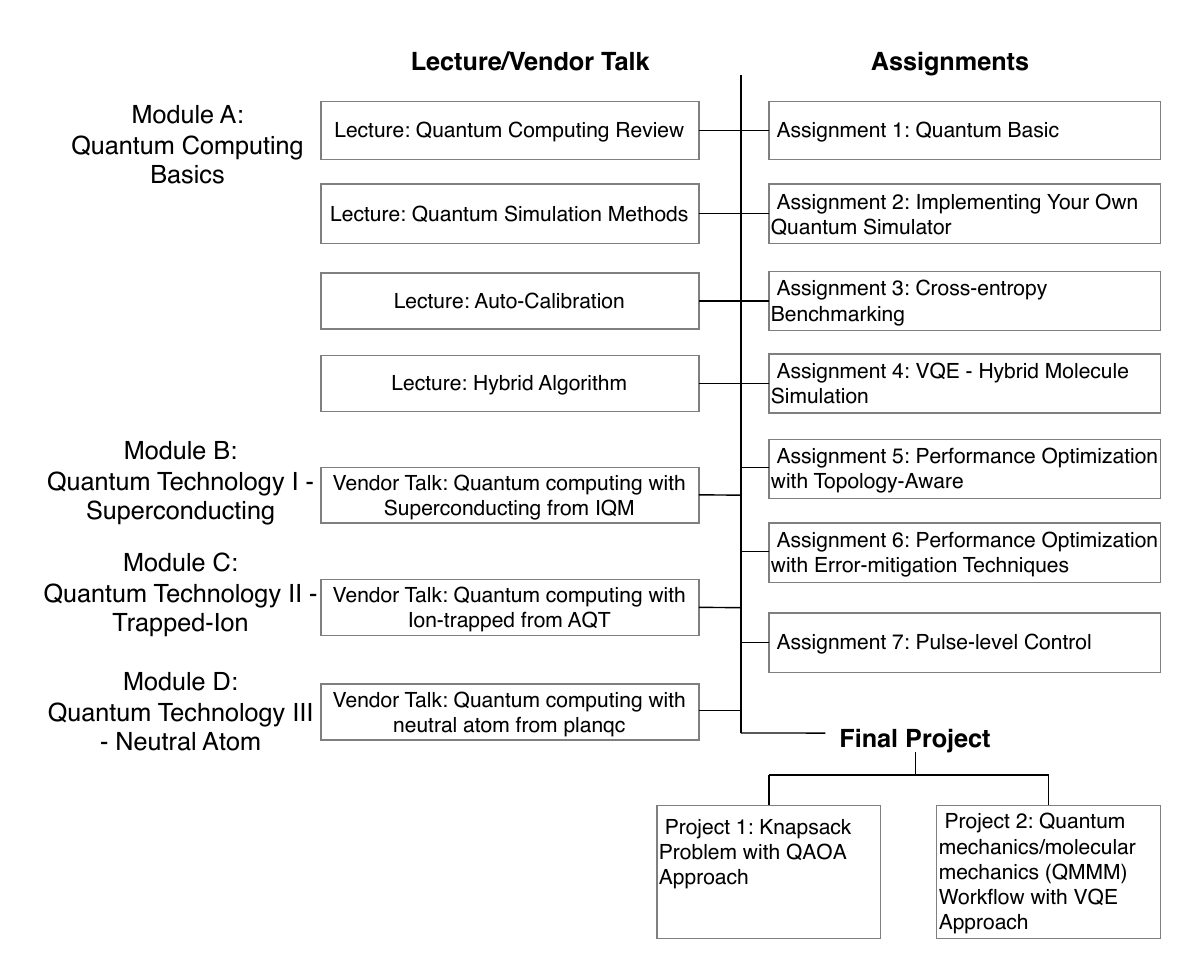}
    \caption{The Q-BEAST practical course with various topics of assignments, lectures/talks, and projects.}
    \label{fig:q-beast_structure}
\end{figure*}

The course is a combination of lectures, vendor talks, structured assignments, and projects. Each assignment is designed to build students' understanding and practical skills progressively. Figure~\ref{fig:q-beast_structure} shows the structure of the Q-BEAST practical course, detailing assignments and projects aligned with the lectures or vendor talks. The course is organized into weekly meetings, each consisting of an assignment and a relevant lecture/vendor talk. For instance, the first week introduces Assignment $1$, which includes basic questions and tasks to review quantum computing principles. There is an overview presentation about the course organization, as well as a lecture reviewing the basics of quantum computing. In the second week, we have a lecture on quantum circuit simulation methods, covering state vectors, density matrices, and matrix product states. Assignment $2$ is introduced next, presenting tasks that require students to build their own quantum simulator using the various simulation methods learned.

The following details each assignment and project shown in Figure~\ref{fig:q-beast_structure}, outlining the topics, objectives, and learning outcomes. In general, our practical course is structured into 4 Modules. Module A, alongside Assignments 1 - 4, focuses on fundamental knowledge and concrete topics in Quantum Computing. Assignments 5 - 7 help students approach techniques to optimize the performance of quantum circuits. In the meantime, we integrate vendor talks specializing in 3 types of quantum technologies, i.e., superconducting, trapped-ion, and neutral atom. Assignments 5-7 and vendor talks comprise the content of Modules B, C, and D. Following the phase of assignments, there is a phase of final projects, which we ask students to implement on a realistic application or a specific optimization problem. Students can flexibly apply optimization techniques to improve performance and compare with the baseline implementation provided.

The specific content of the assignments includes:
\begin{itemize}
    \item \textbf{Quantum computing basics:} We mainly review quantum computing operations, tensor products, and entanglement. There are questions related to the hands-on calculation and implementation tasks that students complete to run on simulators and real devices.
    
    \item \textbf{Implementing a quantum circuit simulator:} Given a source code, students need to complete the implementation of how quantum operations can be simulated. Essentially, they need to verify the principles of state vectors and density matrices to complete the tasks. Then, students follow the process of how quantum operations can be calculated using vectors/matrices to implement the tasks.
    
    \item \textbf{Cross-entropy benchmarking:} The aim is to compare the fidelity of executing quantum circuits between simulators and several kinds of real devices. The cross-entropy benchmarking (XEB) theory refers to the cross-entropy theory~\cite{google2025xeb}.
    
    \item \textbf{VQE -- Hybrid Molecule Simulation:} Students investigate and implement variational principles (in this case, the variational quantum eigensolver, VQE) in quantum computation theory~\cite{tilly2022vqe}. The assignment gives a pre-computed Hamiltonian for the $H_{2}$ molecule. Students implement the ansatz circuit and evaluate the performance of the VQE workflow.
    
    \item \textbf{Performance Optimization with Topology-Aware Qubit Mapping:} Students evaluate the effect of qubit connectivity or QPU topology. For instance, given a GHZ circuit, we can compare the fidelity when running in a naive and a topology-aware way, where the latter specializes in Pareto-optimal qubit selection. The content of this assignment relies on IQM tutorials on superconducting quantum devices \cite{iqm2025tutorials}. We collaborate with IQM for a vendor talk about superconducting technology.
    
    \item \textbf{Performance Optimization with Error-mitigation Techniques:} Students investigate different error mitigation techniques, such as Readout Error Mitigation (REM), Pauli Twirling (PT), and Zero-noise extrapolation (ZNE). Students have the opportunity to perform experiments on real devices to evaluate the noise effect.
    
    \item \textbf{Pulse-level Control:} Students experiment with pulse-level to solve the time evolution of a qubit using a simulator. The assignment helps students familiarize themselves with the concepts of energy relaxation and decoherence of qubits with or without noise. It is aligned with a short presentation on the principle of pulse-level control.
\end{itemize}

For the final projects, there are two use cases: Knapsack problem using the quantum approximate optimization algorithm (QAOA) and hybrid molecular dynamics using combined quantum mechanics and molecular mechanics (QMMM) with the VQE approach. The two topics have different approaches; however, tasks and objectives are the same. 

\begin{itemize}
    \item \textbf{QMMM:} The QMMM concept is introduced by Warshel and Levitt~\cite{warshel1976qmmm}. QM represents quantum mechanics, while MM refers to molecular mechanics models. If only MM is used, the available molecular mechanics force fields do not sufficiently reflect the model processes in which chemical bonds are broken or formed. A full quantum mechanical description can overcome these limitations. The QMMM methods treat a small part of the system at the quantum chemistry level. Additionally, we can retain the computationally cheaper force field for the majority of the system. The use case in this project is a collaboration supported by the research team at the Centre for Computational Science (CCS)\footnote{\url{https://www.ucl.ac.uk/computational-science/}}, University College London (UCL). The CCS is concerned with various aspects of theoretical and computational science, ranging from chemistry and physics to materials science, life sciences, and biomedical sciences.

    \item \textbf{Knapsack:} The Knapsack problem is well known as a classical optimization problem, which has been studied for more than a century~\cite{mathews1896knapsack}. This problem can be defined in quantum computing domains by the value and weight of each item, the total capacity of the knapsack, and the penalty associated with overloading the knapsack. Then, we can formulate the underlying cost Hamiltonian, which is the target to be solved by using the QAOA approach.
\end{itemize}

We provide students with a reference code, which they need to complete and ensure it runs on both simulators as well as real quantum devices. After that, various experiments need to be performed and compared. Students are then asked to consider methods for optimizing performance. The optimization techniques can include topology effects of quantum systems, error mitigation methods, or insights gained from lab assignments to investigate. Finally, students conduct a comprehensive analysis of performance and challenges in scaling up problem sizes to improve speedup.

A major focal point in assignments (in particular, in Assignment $4$) and projects is investigating the benefits of hybrid quantum-classical algorithms. Students are asked to profile the classical parts for potential parallelization while managing the quantum parts using real quantum hardware. For instance, one of the key insights students take away from the QMMM project is the performance bottleneck when relying solely on the quantum hardware for solving variational problems. In a test case using a Variational Quantum Eigensolver (VQE), a single QMMM iteration could take up to $90$ minutes on real quantum hardware. To address this, students explore the concept of Hamiltonian partitioning on the HPC side by decomposing the total Hamiltonian into smaller fragments and sending only the minimal, optimized quantum parts to the quantum backend. This hybrid strategy not only reduces the computational burden of quantum computing but also demonstrates a meaningful way to balance the strengths of HPC and quantum systems.\\

In conjunction with assignments and projects, Q-BEAST organizes lectures and vendor talks. Lectures are presented on key topics related to the tasks of the assignments. Vendor talks offer students the opportunity to listen to speakers from vendors and discuss current quantum computer technologies. Specifically, there are four lectures and three vendor talks.

\begin{itemize}
    \item \textbf{Lectures:} We review quantum computing basics in the first week. The next lecture talk is about quantum simulation methods, divided into weak and strong simulations. The concept revolves around simulation methods utilizing state vectors, density matrices, and matrix product states. Linking with the corresponding assignment, students learn how to implement their quantum simulators. Alongside Assignment $3$ is a lecture about auto-calibration and how it works in quantum computers at LRZ. In the lecture on hybrid algorithms, we introduce the variational principle in classical-quantum computing. The lecture explains how classical optimization problems can be mapped to a quantum domain, how to prepare the Hamiltonian, and how to build the ansatz circuit.

    \item \textbf{Vendor talks:} We collaborate with vendors from different quantum technologies, including IQM\footnote{\url{https://www.meetiqm.com}} for superconducting quantum computers, AQT\footnote{\url{https://www.aqt.eu}} for trapped-ion quantum computers, and planqc\footnote{\url{https://www.planqc.eu}} for neutral atom quantum computers. The vendor's talks mainly cover the background of how a quantum computer is built based on the corresponding technology. Students have opportunities to discuss these emerging technologies.
\end{itemize}

\section{Teaching Support Structure}

Students are assigned to work in groups of two or three. We use Git to manage all student submissions and check their commits. Students can collaborate with team members using Git, where they can also manage each member's contributions in a group. The deadline for each assignment is typically one to two weeks, depending on the workload. After each assignment submission, two or three groups are selected to give a presentation about their experiments, results, and what they have learned. Following that, all attendees can ask questions, give comments, and discuss the details of the assignment concept or performance aspects. For the final project, the deadline is longer, i.e., three weeks, with an intermediate discussion in the middle to exchange feedback with students about their progress. Ultimately, all groups present their experimental results for the selected projects. The overall evaluation relies on various aspects of what students contribute, from assignments to the final project, such as completing tasks, understanding the assignment/project problems, presenting their work, and working in a group. Besides the weekly meetings in class, a Zulip~\cite{zulip2025edu} channel provides students with a platform to actively ask questions and raise issues. Tutors and students can discuss technical problems, issues with systems, and unclear assignment questions/tasks with each other and come up with solutions together.

\section{Evaluation and Feedback}

As the winter semester of $2024$ is the first time we organized the Q-BEAST Praktikum, we restricted the number of participants. We limited the class size to a maximum of $15$ students to allow us to manage the workload and support the students better. This also aligns with the quantum resource allocation slots available for sharing among users. For the course evaluation, we made an effort to gather feedback from students through a survey about their experiences during the Praktikum. While we understand that the survey results may not fully represent all possible perspectives, we still see them as a helpful basis for improving the course and shaping future iterations. In addition, student feedback from the first-time implementation is particularly interesting, especially in the context of emerging topics such as HPCQC integration, where early impressions can guide future directions of HPCQC education.

The feedback comprises inputs from $14$ of $15$ students enrolled in the Q-BEAST Praktikum course, comprising $10$ Master's and $4$ Bachelor's students. Their study programs are from various backgrounds, including computer science, physics, data science, and robotics. Regarding prior exposure, $3/4$ of the students had previously attended courses on quantum computing. Specifically, they had participated in $1$ to $2$ quantum computing-related courses at university, on online platforms, or through self-study via online forums. The remaining students did not have solid basic quantum knowledge.

Regarding general programming skills, most students have more than $3$ years of experience, as almost all students are from computer science-related programs. Half of the students have been proficient in programming for more than $5$ years; however, regarding quantum programming SDKs like \texttt{qiskit}~\cite{qibm2024qiskit}, which is one of the most popular quantum programming frameworks, $12$ out of $14$ students have less than $1$ year of experience. Only $2$ students have $1$ to $3$ years of experience working with \texttt{qiskit}. This survey point helps to partly map the ability to solve assignments and how long students implement the project in Q-BEAST.

\begin{figure*}[t]
    \centering
    \includegraphics[scale=0.425]{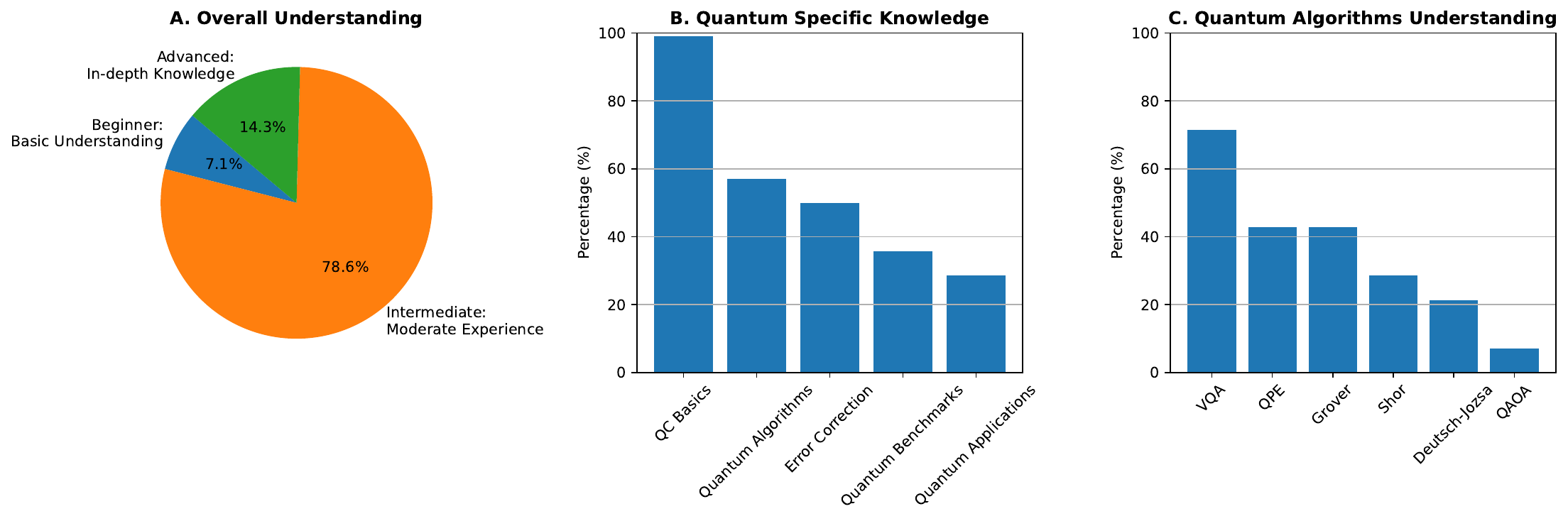}
    \caption{Assessment of Student Knowledge Gain in Quantum Computing through Q-BEAST.}
    \label{fig:q-beast_understanding}
\end{figure*}

The following are the survey results and student feedback about the experiences and outcomes they gained after the course. Figure \ref{fig:q-beast_understanding} (A) shows student self-assessment of quantum computing knowledge and skills, $14.3\%$ reported an advanced level of understanding (in-depth knowledge and hands-on experience), $78.6\%$ reported an intermediate level (moderate experience with practical application), and $7.1\%$ reported still at the beginner level (basic) of understanding. Although almost all the students reported a good understanding of quantum computing basics as shown in Figure~\ref{fig:q-beast_understanding} (B), $57.1\%$ demonstrated familiarity in quantum algorithms, $50\%$ students expressed an understanding of error correction and mitigation followed by relatively lower familiarity in topics related to quantum technologies, such as quantum benchmarking $35.7\%$ and quantum applications $28.6\%$. This data suggests that while students have a foundational understanding of quantum computing fundamentals, they still require additional lectures and guidance on advanced topics, such as benchmarking and realistic applications.

Specifically, regarding the popular quantum algorithms that students are expected to practice around the assignments, Figure \ref{fig:q-beast_understanding}~(C) inquires about students' interests and level of understanding after participating in Q-BEAST. Students expressed varying levels of understanding across different quantum algorithms: $71.4\%$ in variational quantum algorithms, $42.9\%$ in the quantum phase estimation algorithm, $42.9\%$ in Grover's algorithm, $28.6\%$ in Shor's algorithm, $21.4\%$ in Deutsch-Jozsa algorithm, and $7.1\%$ in QAOA.

\begin{figure*}[t]
    \centering
    \includegraphics[scale=0.4]{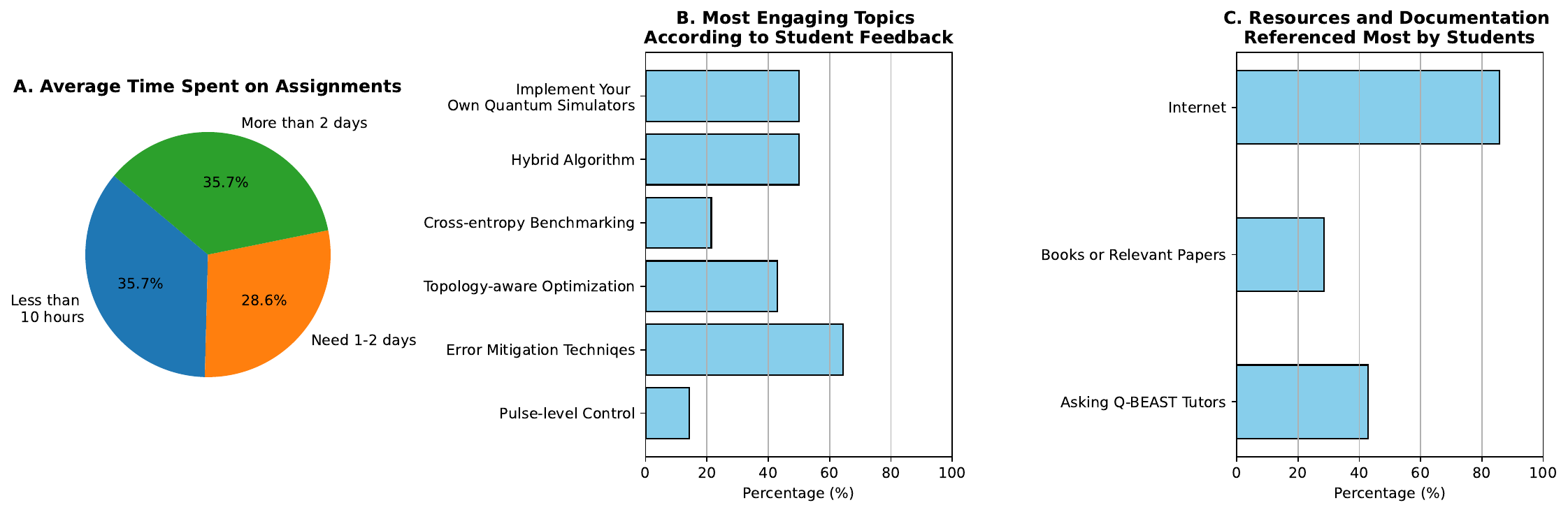}
    \caption{Student Feedback on Time Management, Topic Interest, and Material Reference in Q-BEAST Assignments and Projects.}
    \label{fig:q-beast_experience}
\end{figure*}

In terms of the Q-BEAST Praktikum aspect and experience, the overall satisfaction is positive. As part of the course structure, students collaborated in teams to complete their assignments. The students had to spend a considerable amount of time to complete the assignments: $35.7\%$ needed less than $10$ hours, $28.6\%$ needed $1$-$2$ days, and $35.7\%$ needed more than $2$ days. These values are shown in Figure \ref{fig:q-beast_experience} (A), which indicates that the assignments are manageable and not overly demanding. The majority of the students reported that the assignments helped reinforce their understanding. Additionally, they found the course relevant to their future goals, aligning with their research interests. This suggests that the course structure and content align with most students' aspirations. The difficulty of assignments was rated as moderate, and they are well-balanced; no student found them extremely difficult or very easy. The practical course should maintain a balanced difficulty, with projects focused on various use cases. In addition, Figure \ref{fig:q-beast_experience} (B) shows which assignment topics students found the most interesting. With the given topics in Q-BEAST, the four assignments most cited are as follows:
\begin{itemize}
    \item Implementing your own quantum simulators based on the state vector or density matrix methods;
    \item Hybrid algorithms, i.e., VQE with a small-scale example of molecule simulation;
    \item Topology-aware, based on qubit connectivity and calibration data to optimize the fidelity of quantum circuits.
    \item Error mitigation techniques, such as readout error mitigation (REM), Pauli twirling (PT), and zero-noise extrapolation (ZNE).
\end{itemize}

Shown in Figure \ref{fig:q-beast_structure} is the Q-BEAST structure, lectures and vendor talks are integrated with weekly assignments. Students found lectures on ion-trapped and neutral atom quantum technologies challenging to follow, while the superconducting qubit technology is considered easier to follow. This could suggest they need additional support with supplementary explanations and exercises to track. Besides, when working on the assignment and participating in lectures/vendor talks, Figure \ref{fig:q-beast_experience} (C) shows that the most helpful reference materials are found when students actively search by themselves on the internet; reference content from books or relevant papers accounts for a smaller proportion. Interaction with tutors accounts for half when students encounter system problems or need to discuss and clarify questions and tasks in the assignment.

\begin{figure*}[t]
    \centering
    \includegraphics[scale=0.375]{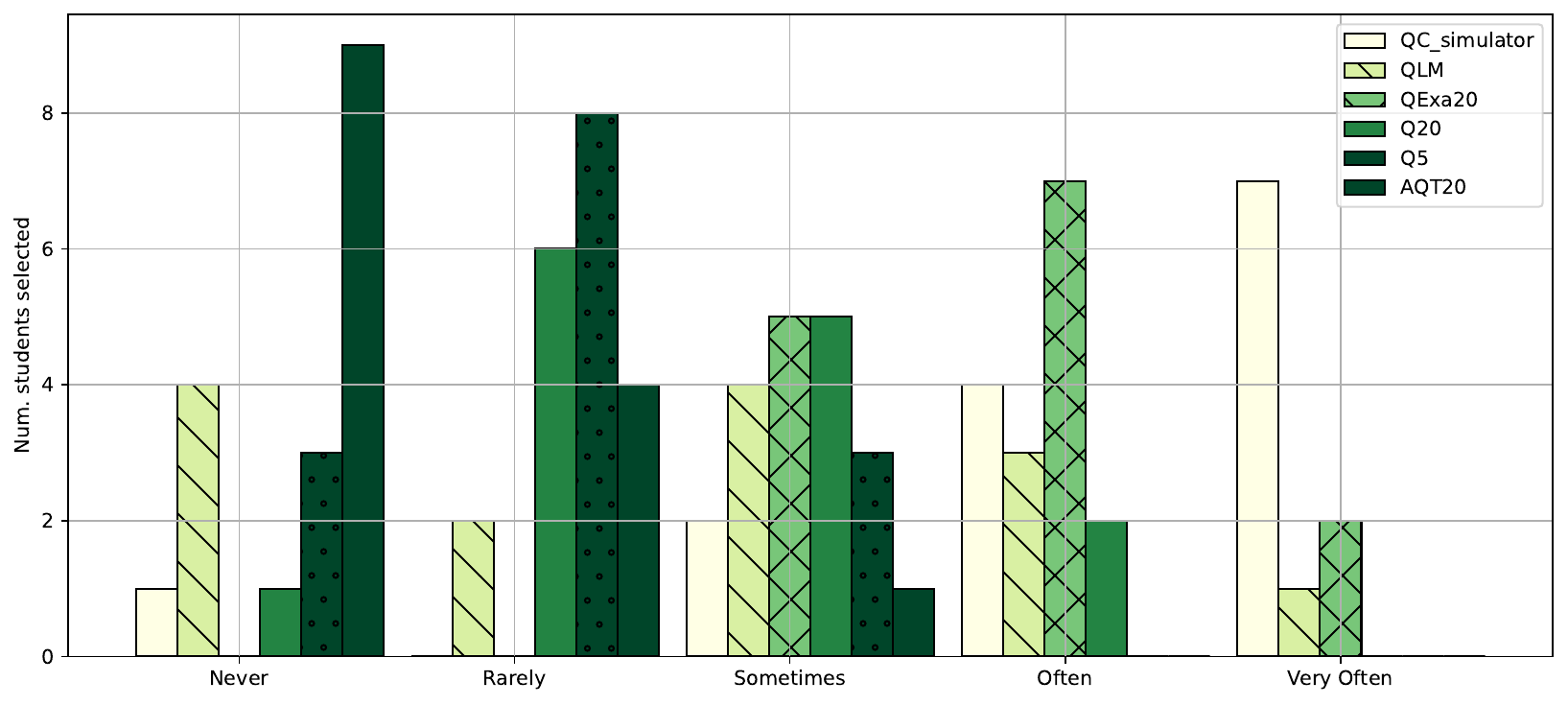}
    \caption{Student Feedback on Quantum Computing System Access and Frequent Resource Usage in Q-BEAST Assignments}
    \label{fig:q-beast_systems_usage}
\end{figure*}

In Q-BEAST, we provide students with two access paths for submitting quantum jobs. As mentioned in the previous sections, MQSS manages the integration between HPC and QC systems. The first access path is via the Munich Quantum Portal, where students are provided accounts to access the portal, create tokens, and submit quantum jobs remotely from their laptops. The second access path is via HPCQC, where students' accounts are allowed to access an HPC system and can submit HPCQC jobs via the SLURM batch job manager. Students can experiment with various quantum computing backends regarding quantum devices or backends. Apart from the most common quantum simulators known as \texttt{qiskit} (fake backends or denoted by QC\_simulator shown in Figure~\ref{fig:q-beast_systems_usage} in general), we have a quantum hardware emulator called QLM (Quantum Learning Machine from Eviden\footnote{\url{https://eviden.com}} known as Qaptiva HPC) capable of state-vector simulations up to $38$ qubits in double precision. There are three superconducting devices from IQM with the codenames Q$5$, Q$20$, and QExa$20$, where Q$5$ and Q$20$ are mainly used for research, having $5$ and $20$ qubits, respectively. The QExa$20$ machine with $20$ qubits is a production system showcasing HPC-QC integration at LRZ. Additionally, there is a trapped-ion device from AQT with the codename AQT$20$ having $12$ qubits. Figure~\ref{fig:q-beast_systems_usage} shows the student feedback on the access to different quantum backends and frequent usage. In the groups of ``often'' and ``very often'', the survey shows that the most frequently used backends are QC\_simulator, QLM, and QExa$20$ because of their availability. The other systems are not used as often, partly due to related construction work, maintenance, software updates, and the busy job queue because quantum systems currently have to be shared between users/research groups.

Based on student feedback, $71.4\%$ of them prefer to use the MQP for submitting quantum jobs with \texttt{mqp-qiskit-provider}\footref{mqssnote}, while $28.6\%$ prefer to use HPCQC. With MQP, the access procedure is quite simple, and students can submit jobs from their laptops. For HPCQC, the primary procedure involves accessing the HPC system, which may require authentication steps depending on the security policy of the computing centers. HPCQC job submission follows the SLURM template, for example, preparing a job script and using the SLURM command to submit the job. MQP may be simple but only suitable for purely quantum jobs or small-scale use cases. HPCQC may be less user-friendly, but it is ideal for large-scale use cases, especially for cases that require HPC-side computational resources coupled with QC systems' results. These points are part of why students prefer MQP over HPCQC; almost all Q-BEAST jobs and experiments are not considered on a large scale.

\begin{figure}[t]
    \centering
    \includegraphics[scale=0.465]{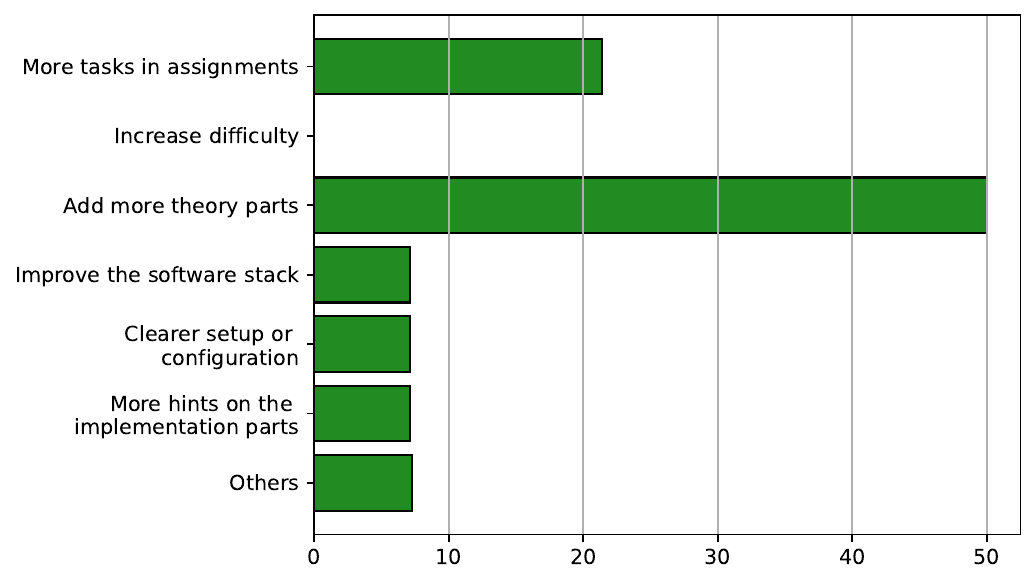}
    \caption{Student feedback on improvement aspects for the quantum practical course based on the Q-BEAST structure.}
    \label{fig:q-beast_improve_aspects}
\end{figure}

Students appreciated several aspects of the course, particularly the emphasis on fundamental concepts, customizing circuits to access real hardware, performing quantum algorithms, and interactive lectures. They also liked the theoretical explanations, error correction techniques, and opportunities to characterize different systems. A key area for improvement is the quality of the lecture handouts and supporting material to facilitate better learning. The assignments and handouts should provide more detailed instructions, additional exercises, and references for the advanced topics. The course assignments strive to strike a balance between theoretical depth and hands-on exercises, helping students gain practical experience. Some students required extra support on setup, library compatibility, and troubleshooting. As Figure~\ref{fig:q-beast_improve_aspects} shows, students expected more tasks, exercises, and theory parts in the assignments. To further improve, they suggested allocating more time for the final projects and providing better explanations for the assignment tasks, in which $\approx 21\%$ voted for more tasks and $50\%$ for more theory parts. Some prefer to gain additional experience in technical setup, such as verifying which libraries work, creating environments, and explaining template codes.

\section{Related Works}

Several initiatives have sought to integrate practical quantum computing education into academic and research institutions. Programs such as Qiskit Global Summer School \cite{qbraid2024qiskitsummerschool}, IBM Quantum Learning \cite{ibm2025qclearning}, or Microsoft’s Quantum Development Kit workshops \cite{microsoft2024devkit} offer theoretical and coding experiences. Primarily, these programs focus on software development and algorithmic principles that effectively introduce quantum computing concepts. However, the experimental and system characterization components are not covered. Besides, there are educational toolkits like QuTiP \cite{johansson2012qutip} and Cirq \cite{google2025cirq} that provide simulation-based platforms for understanding quantum mechanics. Although these toolkits are powerful in scope, they are mainly software-centric, and the terms of HPCQC integration are also not mentioned to leverage real HPCQC hardware benchmarking tasks.

Academic institutions such as MIT's quantum labs also developed quantum computing lab courses \cite{englund2025labqcmit}. ETH Zurich’s quantum center offers various quantum computing lectures that integrate experimental and theoretical learning~\cite{ethzurich2025qcclasses}. 
Q-BEAST differentiates itself by combining HPC infrastructure, different quantum technologies, and a structured curriculum centered around benchmarking and performance evaluation. Its close alignment with the Munich Quantum Software Stack (MQSS)~\cite{schulz2023mqss} and integration with the BEAST infrastructure enables students to explore quantum technologies and understand the challenges of deploying as well as optimizing quantum workflows in hybrid HPCQC environments. The teaching approach in Q-BEAST is inspired by a HPC practical course for modern parallel computing architectures, namely BEAST-Lab~\cite{raoofy2024beastlab}, which we have instructed since 2020. Q-BEAST offers a unique perspective on the landscape of quantum computing education by emphasizing practical, system-level analysis.

\section{Lessons Learned, Challenges, and Outlook}

Throughout the deployment of the Q-BEAST Praktikum, we have gathered valuable feedback from students, which has helped identify key areas for future development and improvement. These insights can enhance the course's effectiveness and suggest the ongoing evolution of the support infrastructure and software tools. The following are the key lessons learned:

\begin{itemize}
    \item One of the significant observations was the importance of up-to-date calibration data and clear indications of system workload. More helpful information about the live system status would allow students to perform experiments and analyze the performance effectively.

    \item Technical setup and HPC-QC system access are another recurring theme. Students expressed the need for smoother and simpler procedures. Particularly via an HPC system with SSH and authentication steps, users/students need to become familiar with it. Slow remote access can also cause confusing experiences for experiments with short execution times but a large number of jobs.

    \item Functionality in job management was another area of concern. Students suggested returning a queue object for jobs and runtimes, along with the ability to view all queued jobs. These features would provide better feedback and transparency when working with remote resources.

    \item On the software side, there were issues like missing essential tools, libraries, or conflicting software packages. Guiding students through technical issues and configuring experimental environments will help reduce errors that can sometimes be time-consuming while working on assignments.

    \item Broader integration possibilities, such as more quantum providers, were also highlighted. This would open up a wider range of learning opportunities and system comparisons.
    
    \item Students advocated for more flexible scripting support and more extensive API documentation. These improvements would better enable the adaptation of tools for diverse experiments.
    
    \item Finally, adding more functionality and system status information to the MQP's dashboard will help reduce confusion and streamline students' workflow for performing experiments, such as queue information of jobs, statistics of used computation time, the status of qubits, etc.
\end{itemize}

Based on these lessons, future iterations of the Q-BEAST Praktikum will prioritize improvements to documentation, user accessibility, resource integration, and system information/status. The feedback has underscored the value of responsive infrastructure and tools that are essential for productive hands-on quantum computing education. Ongoing projects within the Munich Quantum Valley initiative, as well as the further development of MQSS, will play a central role in these improvements. The MQSS is extending the programming interfaces it supports, such as Pennylane~\cite{bergholm2022pennyautohybrid} and CUDA-Q~\cite{kim2023cudaq}, that could be named ``MQSS Pennylane Adapter'' and ``MQSS CUDAQ Adapter''. Besides, our goal is to foster an HPCQC environment that encourages students to become active contributors between supercomputing centers and academic institutions.

\section{Conclusion}

The Q-BEAST Praktikum represents a significant step toward bridging the gap between theoretical quantum computing education and practical/hands-on experience with actual quantum and HPC systems. By integrating diverse technological components -- from HPC platforms to quantum hardware and software stacks -- Q-BEAST provides students with an environment for exploring the experimental aspects of quantum computing.

Students are exposed to the interdisciplinary nature of modern quantum research through structured assignments, collaborative projects, and tools from the Munich Quantum Software Stack (MQSS). The course has revealed practical insights into system usability, software limitations, and areas for further educational support. These enhancements will make future course iterations more accessible, flexible, and aligned with the real-world demands of quantum computing. The continued evolution of quantum technologies and the expansion of educational infrastructures, such as Q-BEAST, will be essential in preparing the next generation of engineers and researchers.

\section*{Acknowledgment}

This work was supported by the German Federal Ministry of Research, Technology and Space (BMFTR) and the Bavarian State Ministry of Science and the Arts through funding as part of the Munich Quantum Valley (MQV) and Bavarian High-Tech Agenda, especially MuniQC-Atoms and Q-DESSI. The authors gratefully acknowledge the Leibniz Supercomputing Centre (LRZ) for providing computation time and access to HPC systems that were essential for the practical execution of the Q-BEAST course.

We would like to express our sincere thanks to Stefan Pogorzalek and Stefan Seegerer (IQM), Christian Sommer and Felix Rohde (AQT), and Adrian Vetter (planqc) for their valuable collaboration and insightful talks on their respective quantum technologies. Their contributions significantly enriched the course content and student engagement. We also extend our appreciation to Thomas Bickley, Angus Mingare, and Peter Coveney for their collaboration on the QMMM project, from which elements were successfully adapted into a student project within the Q-BEAST structure. Finally, we thank all the students and participants of the Q-BEAST Praktikum for their enthusiasm, feedback, and active involvement, which helped ensure the course ran smoothly.

\bibliographystyle{IEEEtran}
\bibliography{references}

\vspace{12pt}

\end{document}